\documentclass[12pt,a4paper,notitlepage,fleqn]{article}

\usepackage[verbose]{geometry}
\usepackage[symbol]{footmisc}
\usepackage[T1]{fontenc}
\usepackage{authblk}

\makeatletter

\newcommand{\fboxsubsec}[1]{
	\begin{flushleft}
		#1
	\end{flushleft}
	}
\renewcommand{\subsection}{\@startsection{subsection}{2}{0pt}
	{1ex}
	{0.5ex}
	{\reset@font\it\fboxsubsec}
	}
\makeatother

\title{On the Evolution of Cryptocurrency Market Efficiency}%

\author{Akihiko Noda$^{a,b}$\thanks{\scriptsize Corresponding Author. E-mail: noda@cc.kyoto-su.ac.jp, Tel. +81-75-705-1510, Fax. +81-75-705-3227.}

{\scriptsize ${}^{a}$ \it Faculty of Economics, Kyoto Sangyo University, Motoyama, Kamigamo, Kita-ku, Kyoto 603-8555, Japan}

{\scriptsize ${}^{b}$ \it Keio Economic Observatory, Keio University, 2-15-45 Mita, Minato-ku, Tokyo 108-8345, Japan}}

\date{This Version: \today}

\renewcommand\thefootnote{\arabic{footnote}}

\usepackage{graphicx}

\usepackage[]{natbib}%
\usepackage{amsmath,amssymb}%
\usepackage{ascmac}%
\usepackage{multirow}%
\usepackage{lscape}%
\usepackage{subfigmat}

\usepackage{pifont}%
\usepackage{arydshln}%
\usepackage[format=hang]{caption}
\usepackage[all]{xy}
\usepackage{url}
\bibpunct{(}{)}{;}{a}{}{,}

\def\hsymbu#1{\smash{\lower1.7ex\hbox{\huge$#1$}}}

\newcommand{\citetapos}[1]{\citeauthor{#1}'s \citeyearpar{#1}}

\begin{document}

\begin{titlepage}

\renewcommand{\thepage}{}
\renewcommand{\thefootnote}{\fnsymbol{footnote}}

\maketitle

\vspace{-10mm}

\noindent
\hrulefill

\noindent
{\bfseries Abstract:} This study examines whether the efficiency of cryptocurrency markets (Bitcoin and Ethereum) evolve over time based on \citetapos{lo2004amh} adaptive market hypothesis (AMH). In particular, we measure the degree of market efficiency using a generalized least squares-based time-varying model that does not depend on sample size, unlike previous studies that used conventional methods. The empirical results show that (1) the degree of market efficiency varies with time in the markets, (2) Bitcoin's market efficiency level is higher than that of Ethereum over most periods, and (3) a market with high market liquidity has been evolving. We conclude that the results support the AMH for the most established cryptocurrency market.\\



\noindent
\hrulefill

\end{titlepage}

\bibliographystyle{asa}


\section{Introduction}\label{sec:crypto_intro}

Since \citetapos{nakamoto2008btc} description of a digital cryptocurrency named Bitcoin, cryptocurrency markets have expanded, and their total market capitalization reached USD 800 billion by January 2018. However, these markets have subsequently experienced a crisis, and their total market capitalization decreased to USD 100 billion by the end of 2018.\footnote{The historical data for total market capitalization are available at the web page of CoinMarketCap (\url{https://coinmarketcap.com/charts/}).} As such, the changes in market capitalization suggest that investors treat cryptocurrencies as an asset -- but this does not necessarily mean they do not also treat it as, say, a currency. Further, economists consider investigating the efficiency of the cryptocurrency market in the sense of \citet{fama1970ecm} to be essential for evaluating the price mechanism of financial markets. Therefore, several recent studies on cryptocurrency markets have aimed to determine whether these markets are efficient.

There exists a large body of literature on the weak-form of \citetapos{fama1970ecm} efficient market hypothesis (EMH) for cryptocurrency markets, especially the Bitcoin market.\footnote{As described in \citet[p. 739]{malkiel1992emh}, markets are said to be efficient in the weak-form sense if the information set only includes the history of prices or returns.} However, the market efficiency of cryptocurrency has been a subject of controversy between the proponents and opponents of the EMH. For example, \citet{urquhart2016tib}, \citet{nadarajah2017oib}, \citet{bariviera2017tib}, \citet{khuntia2018amh}, \citet{kristoufek2018obm}, \citet{tiwari2018ieb}, and \citet{dimitrova2019scb} conclude that the Bitcoin market is almost efficient. In contrast, \citet{yonghong2018tvl}, \citet{cheah2018lmi}, \citet{ai2018eml}, and \citet{vidai2019wec} present empirical results that do not support the EMH for this market. We suspect that one of the reasons for this controversy is that market efficiency varies over time. In this context, \citet{lo2004amh} proposes the adaptive market hypothesis (AMH) as an evolutionary alternative to the EMH, reinforcing the view that the market evolves over time, as does market efficiency. The most important implication of the AMH is that market efficiency can arise from time to time due to changing market conditions such as behavioral bias, structural change, and external events. Specifically, Lo estimates the time-varying first-order autocorrelation of returns on the U.S. stock market using 60-month rolling windows and shows that stock market efficiency continuously evolves over time. His empirical results suggest that the AMH may be supported by not only the stock market but also other financial markets.

To examine the AMH, two approaches have been adopted in the literature. The first is based on a conventional statistical test to examine the AMH under the split samples or the rolling-window method. In practice, \citet{urquhart2016tib}, \citet{nadarajah2017oib}, \citet{khuntia2018amh}, \citet{kristoufek2018obm}, \citet{chu2019amh}, \citet{dimitrova2019scb}, and \citet{vidai2019wec} employ conventional statistical tests under the split samples to examine the AMH for the Bitcoin market. In particular, \citet{khuntia2018amh} and \citet{chu2019amh} are related to this study because they employ a family of \citetapos{dominguez2003tmd} test statistics to explore whether the Bitcoin price follows the martingale difference sequences. \citet{khuntia2018amh} show the time-varying return predictability in the Bitcoin market using a family of \citetapos{dominguez2003tmd} test statistics in a rolling-window framework and conclude that market efficiency evolves with time and validates the AMH in bitcoin market. \citet{chu2019amh} test the AMH in a similar manner to \citet{khuntia2018amh} using high-frequency data and find that the AMH is supported in the Bitcoin market.

However, these methods have the underlying empirical problem of choosing an optimal window width for the test statistics. Unlike these methods, a GLS-based time-varying model, which is the second approach to examining the AMH, has the superior property that it does not depend on sample size. In this approach, the degree of market efficiency is measured together with its statistical inference. Some studies employ a generalized least squares (GLS)-based time-varying model to estimate the degree of market efficiency on the international stock markets.\footnote{See \citet{ito2014ism,ito2016eme} and \citet{noda2016amh} for details.} \citet{noda2016amh} tests the AMH using Japanese stock market data and concludes that the degree of market efficiency varies with time.

As such, this study examines the AMH on cryptocurrency markets from the viewpoint of market efficiency based on two representative cryptocurrencies, Bitcoin and Ethereum. We choose these currencies because their market capitalization accounts for a large portion of the total market capitalization in the cryptocurrency markets. We first estimate the degree of market efficiency using the GLS-based time-varying model with statistical inferences. Second, we analyze the changes in their degrees of market efficiency over time and whether they show different efficiencies depending on trading volume and market capitalization. Finally, we explore what types of markets support the AMH based on the degree of market efficiency, independent of sample size.

This paper is organized as follows. Section~2 presents our method to study market efficiency varying over time based on a GLS-based time-varying model of \citet{ito2014ism,ito2016eme,ito2017aae}. Section~3 describes the daily prices of cryptocurrencies (Bitcoin and Ethereum) to calculate the returns and presents preliminary unit root test results. Section~4 shows our empirical results and discusses whether the AMH is supported in the cryptocurrency markets from the viewpoint of time-varying market efficiency. Section~5 concludes the article.

\section{Method}\label{sec:crypto_model}

\subsection{GLS-Based Time-Varying AR Model}

We employ a GLS-based time-varying autoregressive (TV-AR) model from \citet{ito2014ism,ito2016eme,ito2017aae} to analyze financial data whose data-generating process is time-varying. The conventional AR model,
\begin{equation*}
x_t = \alpha_0 + \alpha_1 x_{t-1} + \cdots + \alpha_q x_{t-q} + u_t,
\end{equation*}
has been frequently used to analyze the time series of the returns of assets, where $\{u_t\}$ satisfies $E[u_t]=0$, $E[u^2_t]=0$, and $E[u_t u_{t-m}]=0 \ \mbox{for all} \ m$. While $\alpha_\ell$'s are assumed to be constant in standard time series analyses, we assume that the coefficients of the AR model change over time. We thus apply a GLS-based TV-AR model to analyze cryptocurrency markets because financial markets have been facing structural changes for several reasons, such as economic/political crises.\footnote{See Lim and Brooks \citet{lim2011esm} for details.}

A GLS-based TV-AR model is expressed as follows:
\begin{equation}
x_t = \alpha_{0,t} + \alpha_{1,t} x_{t-1} + \cdots + \alpha_{q,t} x_{t-q} + u_t, \label{obseq}
\end{equation}
where $\{u_t\}$ satisfies $E[u_t]=0$, $E[u^2_t]=0$, and $E[u_t u_{t-m}]=0 \ \mbox{for all} \ m$. Furthermore, we assume that parameter dynamics restrict the parameters when we estimate a GLS-based TV-AR model using such data, in particular,
\begin{equation}
\alpha_{\ell,t} = \alpha_{\ell,t-1} + v_{\ell,t}, \ (\ell=1,2,\cdots,q), \label{steq}
\end{equation}
where $\{v_{\ell,t}\}$ satisfies $E[v_{\ell,t}]=0$, $E[v^2_{\ell,t}]=0$
and $E[v_{\ell,t} v_{\ell,t-m}]=0 \ \mbox{for all} \ m \ \mbox{and} \ \ell$. We solve a system of simultaneous equations using Equations (\ref{obseq}) and (\ref{steq}).

According to \citet{ito2014ism,ito2016eme,ito2017aae}, a GLS-based TV-AR model has two major advantages over the conventional Bayesian method (e.g., Kalman filtering and smoothing). First, this method is fairly simple and the calculation is fast. Unlike the conventional Bayesian method, no iteration by Markov chain Monte Carlo (MCMC) algorithms is required. Second, prior distributions of parameters are unnecessary when we employ a GLS-based TV-AR model. We can thus employ conventional statistical inferences (e.g., residual-based bootstrap method) on the time-varying estimates to conduct statistical inferences.

\subsection{Time-Varying Degree of Market Efficiency}
We first calculate the time-varying impulse responses from TV-AR coefficients over each period. We then calculate the confidence intervals for each coefficient based on the estimated covariance matrix. While the concept of a GLS-based TV-AR model is quite simple, two caveats exist: (1) a GLS-based TV-AR model is only an approximation of the real data-generating process, which is supposed to be a complex nonstationary process; and (2) we assume the estimated stationary $\mbox{AR}(q)$ model index by period $t$, which is stationary, as a local approximation of the underlying complex process.

We define the time-varying degree of market efficiency based on \citet{ito2014ism,ito2016eme} as follows:
\begin{equation}
 \zeta_t=\left|\frac{\sum_{j=1}^p\hat{\alpha}_{j,t}}{1-\left(\sum_{j=1}^p\hat{\alpha}_{j,t}\right)}\right|.\label{degree}
\end{equation}
We measure the deviation from the zero coefficients on the corresponding time-varying moving-average model to the TV-AR model. Hence, this implies that large deviations of $\zeta_t$ from zero are evidence of market inefficiency. We know that that degree $\zeta_t$ crucially depends on sampling errors. Thus, we construct confidence intervals for $\zeta_t$'s on the condition that the market is efficient. We find the market at time $t$ period is inefficient when $\zeta_t$ exceeds the upper limit at the $t$ period of the intervals. 

Specifically, the interval is constructed as follows. We first identify the returns with the residuals of a TV-AR($q$) estimation under the above hypothesis that all coefficients are zero. Second, $\mathcal{N}$ samples are extracted as an empirical distribution of the residuals. Third, we fit a TV-AR model to the $\mathcal{N}$ bootstrap samples and derive $\mathcal{N}$ sets of estimates. The $\mathcal{N}$ bootstrap samples of $\zeta_t$ are then computed from the estimates. Finally, we construct confidence intervals from the $\mathcal{N}$ bootstrap samples. Therefore, the bootstrap is conducted under the null hypothesis of zero autocorrelation. The estimates of the degree of efficiency exceed the 99\% confidence intervals, which implies a rejection of the null hypothesis of no return autocorrelation at the 1\% significance level.

\section{Data}\label{sec:crypto_data}

We utilize the daily prices of Bitcoin (BTC) and Ethereum (ETH) obtained from the CoinMarketCap website (\url{https://coinmarketcap.com}). The datasets of the two cryptocurrencies have different start dates: April 28, 2013, for BTC and August 7, 2015, for ETH. On the other hand, the end dates are the same for both cryptocurrencies (September 30, 2019). We take the log first difference of the time series of prices to obtain the returns of the cryptocurrencies.
\begin{center}
(Table \ref{crypto_table1} around here)
\end{center}
Table~\ref{crypto_table1} shows the descriptive analysis for the returns. We confirm that the standard deviation of returns on the BTC is lower than those of ETH. This means that the BTC is a more established and unrisked market than ETH because a lower standard deviation of returns indicates better liquidity. For estimations, all variables that appear in the moment conditions should be stationary. We apply the augmented Dickey--Fuller (ADF) test to confirm whether the variables satisfy the stationarity condition. The ADF test rejects the null hypothesis that the variables (all returns) contain a unit root at the 1\% significance level.

\section{Empirical Results}\label{sec:crypto_emp}

We apply the GLS-based TV-AR model from \citet{ito2014ism,ito2016eme,ito2017aae} to obtain the degree of market efficiency. Note that we employ the Bayesian information criterion to select the optimal lag order for the stationary AR($q$) model. Consequently, we choose the AR(6) model for both cryptocurrencies. We measure the cryptocurrency markets' deviation from the efficient condition using Equation (\ref{degree}) because the degree is based on the spectral norm. The degree of market efficiency thus indicates how the market is different from an efficient market. If $\zeta_t=0$ for time $t$, the market is shown to be efficient at that time.

\begin{figure}[!hbp]
 \caption{Time-Varying Degree of Market Efficiency}
 \label{crypto_fig1}
 \begin{center}
 \includegraphics[scale=0.4]{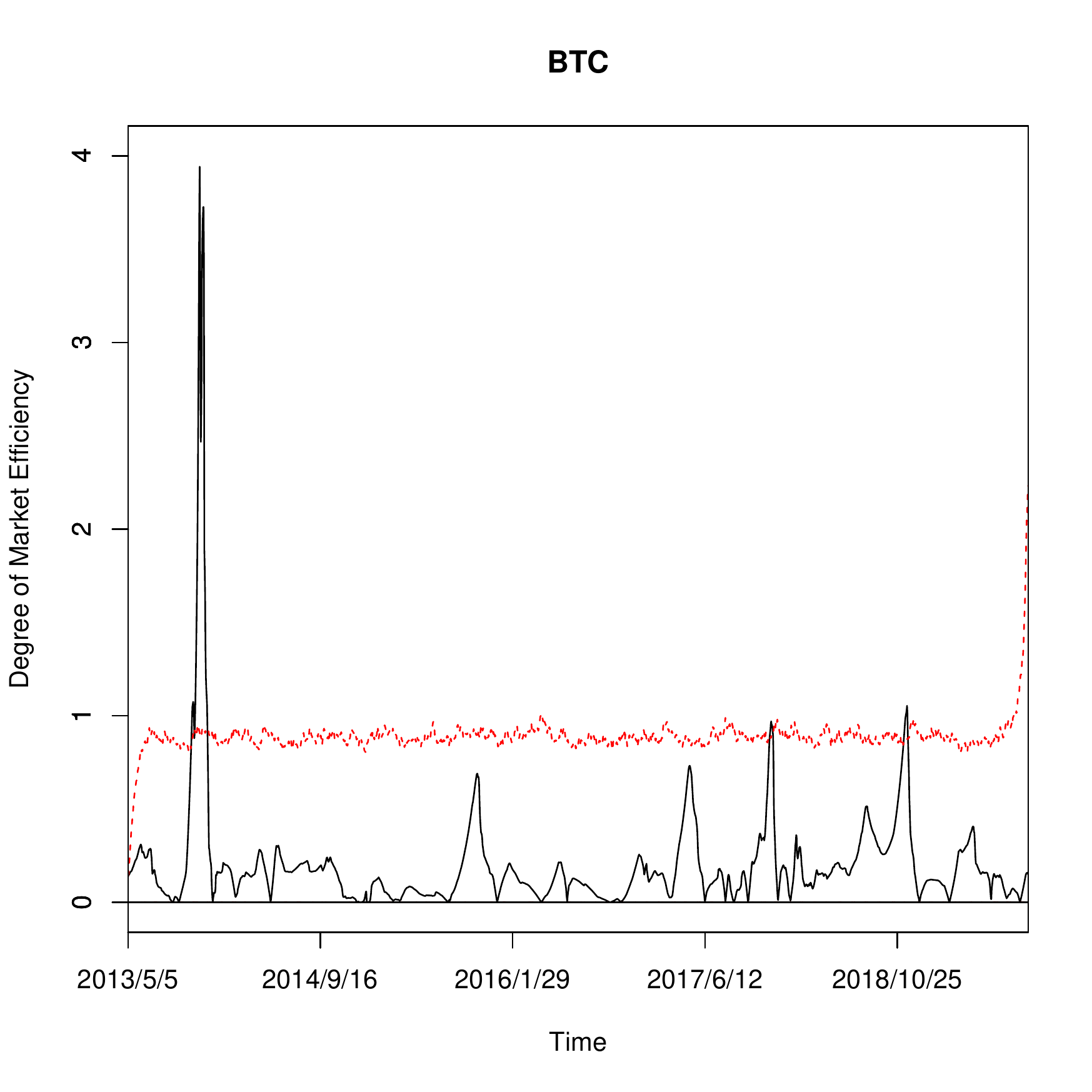}
 \includegraphics[scale=0.4]{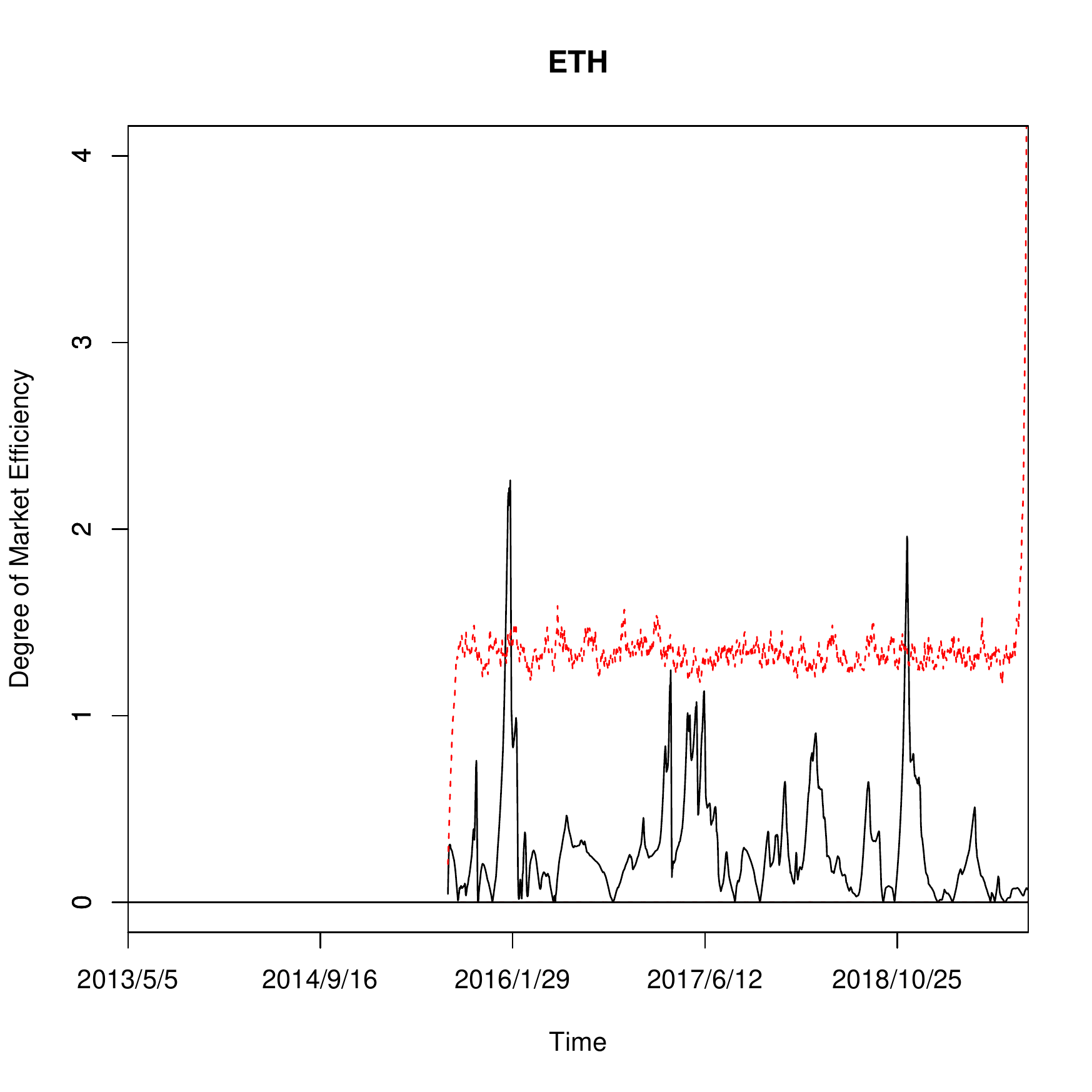}
\vspace*{5pt}
{
\begin{minipage}{420pt}
\scriptsize
\underline{Notes}:
\begin{itemize}
 \item[(1)] The panels of the figure show the time-varying degree of market efficiency for BTC (left panel) and ETH (right panel). 
 \item[(2)] The dashed red lines represent the 99\% confidence intervals of the efficient market degrees.
 \item[(3)] We run bootstrap sampling 10,000 times to calculate the confidence intervals.
 \item[(4)] R version 3.6.3 was used to compute the estimates.
\end{itemize}
\end{minipage}}%
\end{center}
\end{figure}

Figure \ref{crypto_fig1} indicates the degree of market efficiency based on the above TV-AR models. We first find that the degrees of the BTC and ETH change over time. Figure~\ref{crypto_fig1} also demonstrates that the markets are inefficient during some crash or crisis periods. In practice, these correspond with the rapid price decreases of cryptocurrencies (December 2017 and November 2018) and a financial crisis due to ``Mt. Gox'' from November 2013 to February 2014.
\begin{center}
(Figure \ref{crypto_fig1} around here)
\end{center}
We confirm three significant differences among the cryptocurrencies in terms of their degrees of market efficiency. First, since August 14, 2015, BTC has been the most efficient cryptocurrency in the sense of the degree of market efficiency $\zeta_t$, being followed by ETH. The average $\zeta_t$ of BTC and ETH are 0.19 and 0.30, respectively. We also compare the $\zeta_t$s over the same sample period because the periods are different between currencies. The average $\zeta_t$ of BTC is 0.20 using the entire sample for reference. ETH's $\zeta_t$ fluctuates more widely than that of BTC. In fact, the standard deviations of the $\zeta_t$s of BTC and ETH are 0.18 and 0.32, respectively. Third, BTC's $\zeta_t$ was less volatile since the financial crisis due to ``Mt. Gox'' from November 2013 to February 2014, but that of ETH was not.

\begin{figure}[!htbp]
 \caption{Trading Volumes and Market Capitalizations}
 \label{crypto_fig2}
 \begin{center}
 \includegraphics[scale=0.4]{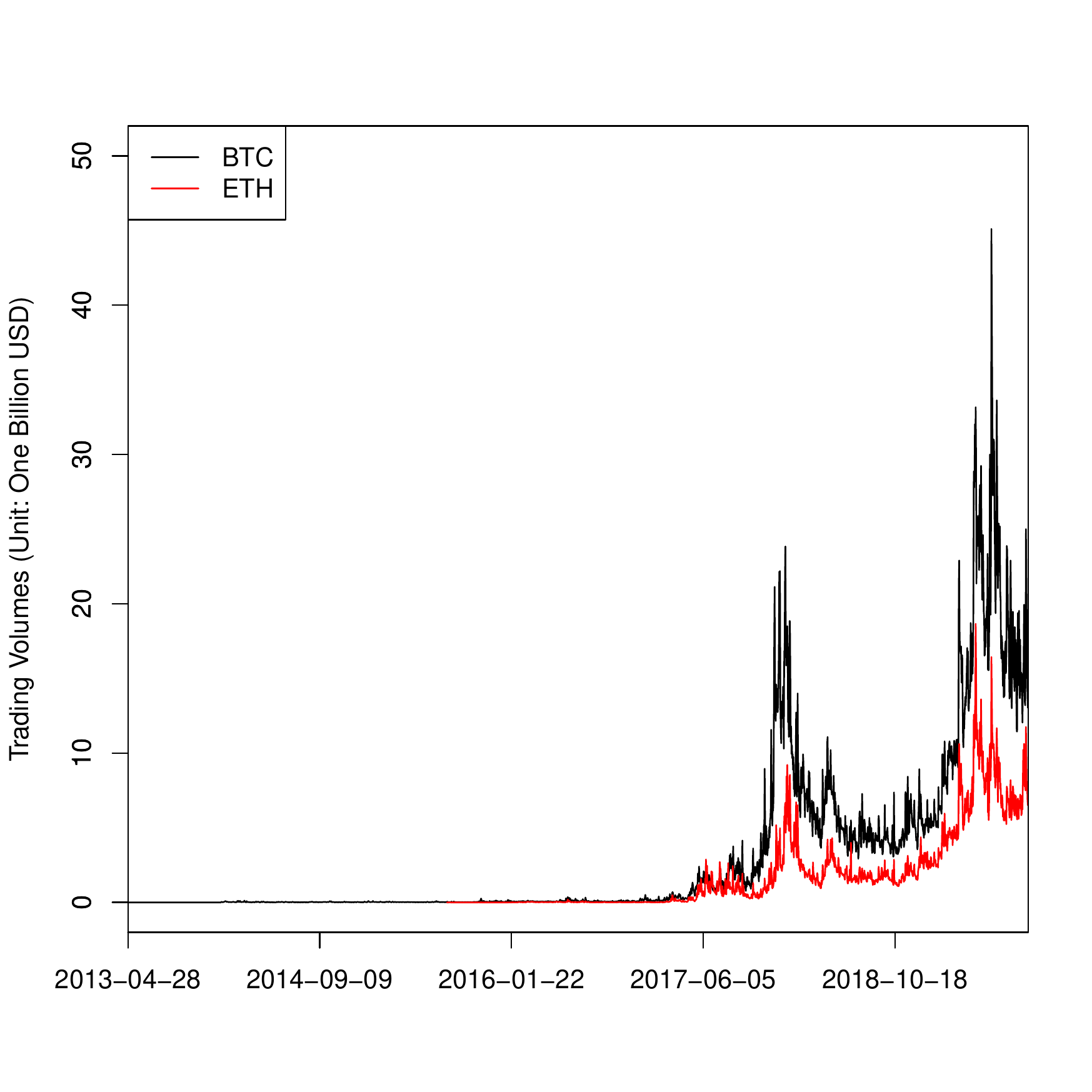}
 \includegraphics[scale=0.4]{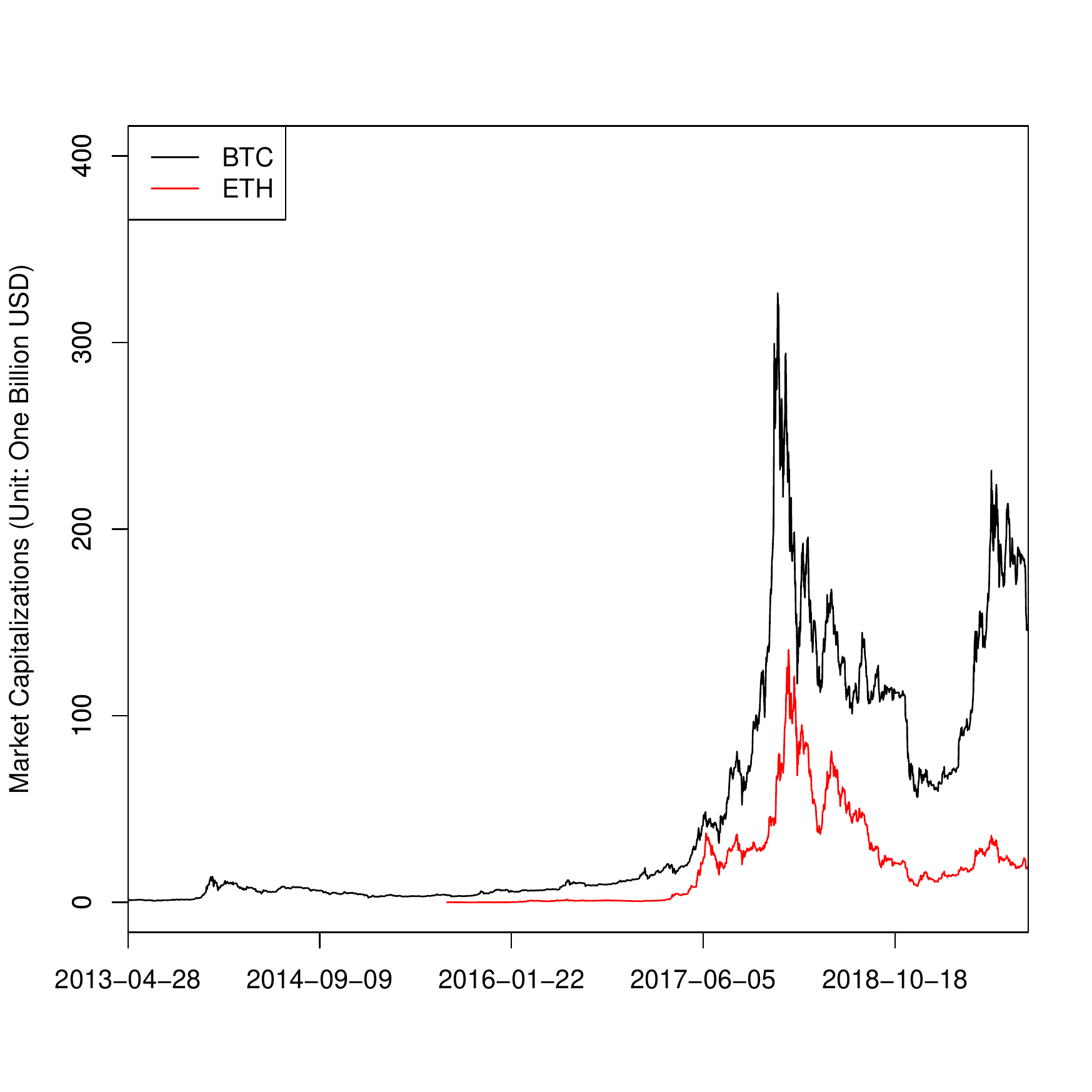}
\vspace*{5pt}
{
\begin{minipage}{420pt}
\scriptsize
\underline{Notes}:
\begin{itemize}
 \item[(1)] The panels of the figure show trading volumes (left panel) and market capitalizations (right panel) for BTC and ETH.
 \item[(2)] The dataset is obtained from the web page of CoinMarketCap (\url{https://coinmarketcap.com/}).
 \item[(3)] R version 3.6.3 was used to compute the statistics.
\end{itemize}
\end{minipage}}%
 \end{center}
\end{figure}

\begin{figure}[!htbp]
 \caption{Percentage of Total Market Capitalization (Dominance)}
 \label{crypto_fig3}
 \begin{center}
 \includegraphics[scale=0.4]{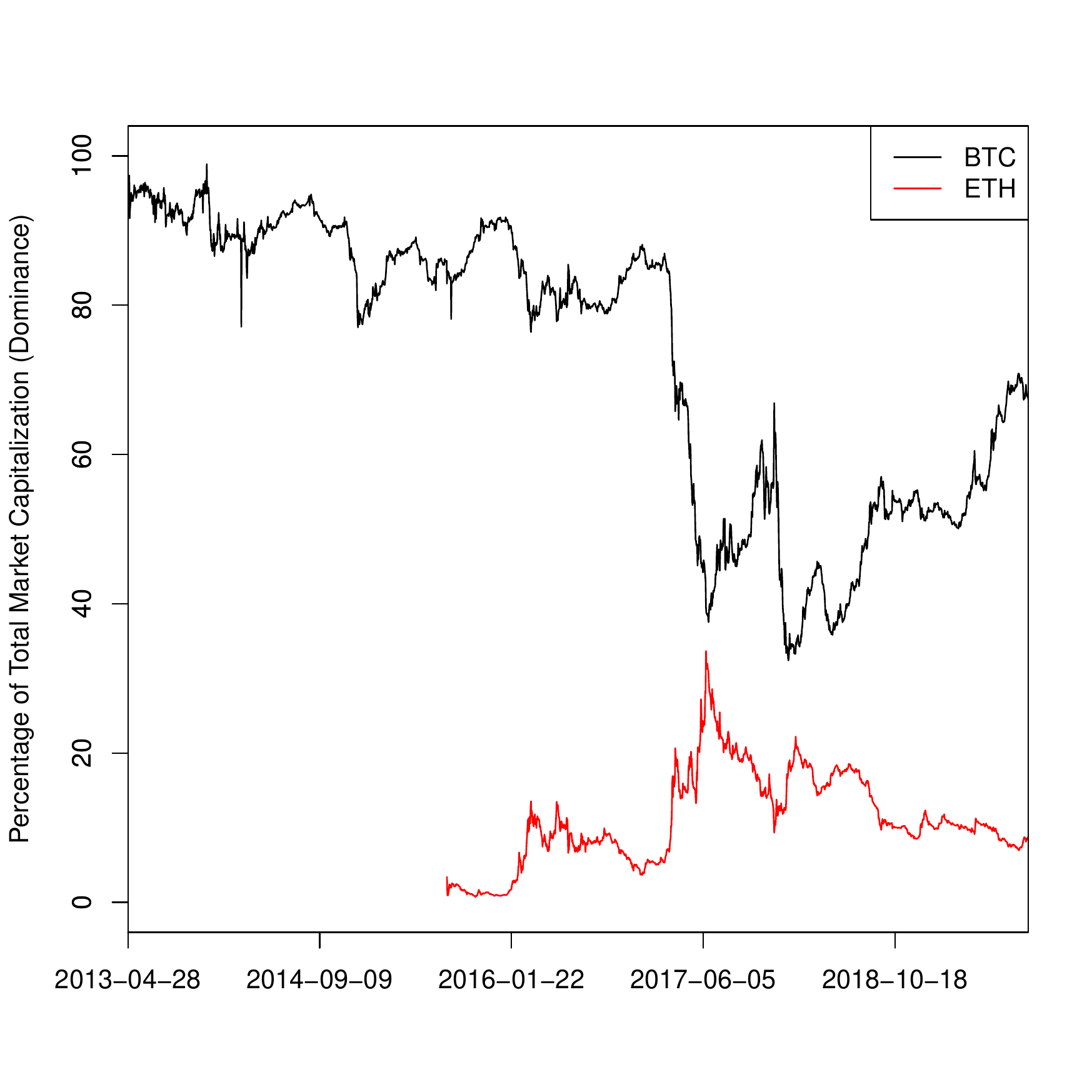}
\vspace*{5pt}
{
\begin{minipage}{420pt}
\scriptsize
\underline{Notes}:
\begin{itemize}
 \item[(1)] The dataset is obtained from the web page of CoinMarketCap (\url{https://coinmarketcap.com/}).
 \item[(2)] R version 3.6.3 was used to compute the statistics.
\end{itemize}
\end{minipage}}%
 \end{center}
\end{figure}

The differences between the BTC and ETH in terms of trading volumes, market capitalization, and percentage of total market capitalization (dominance) might explain these differences in the degree of market efficiency $\zeta_t$, as shown in \citet{brauneis2018pdc}, \citet{wei2018lme}, and \citet{khuntia2020alm}.
\begin{center}
(Figures \ref{crypto_fig2} and \ref{crypto_fig3} around here)
\end{center}
Figure~\ref{crypto_fig2} demonstrates that trading volumes and market capitalizations are quite different between BTC and ETH. Additionally, it is widely known that the market capitalization of BTC and ETH accounts for a large portion of the total market capitalization of the cryptocurrency market. Figure~\ref{crypto_fig3} indicates the changes in the percentage of total market capitalization (dominance) for BTC and ETH. We confirm that the degree of market efficiency of BTC and ETH declines when the dominance changes drastically (early 2017, late 2017, and late 2018). This means that the dominance and trade openness among cryptocurrencies may affect the market efficiency. Empirically, \citet{khuntia2020alm} confirm the time-varying or adaptive behavior of long memory in the volatility of Bitcoin returns and conclude that the long memory of the volatility of returns can be explained by trading volume.\footnote{In a different context, \citet{lim2011toi} and \citet{noda2016amh} show that trade openness is associated with the market efficiency of stock markets.}

Moreover, the empirical results are consistent with \citet{urquhart2016tib}, who shows that market efficiency improves after late 2013 when using sub-sample estimation. In particular, we find that the degree of market efficiency of BTC sometimes declines relatively (e.g., late 2015, early 2017, and early 2018), but it does not achieve the level of an absolutely inefficient market with the exception of a period of rapid price decrease in late 2018. Conversely, that of ETH fluctuates widely and often reaches the level of absolute inefficiency (e.g., early 2016, mid-2017, and late 2018). This implies that the BTC market reflects shock, whereas the ETH market does not. Thus, the empirical results support the AMH on the more qualified cryptocurrency market as shown in \citet{khuntia2018amh} and \citet{chu2019amh}, which examine the AMH on the Bitcoin market.

\section{Concluding Remarks}\label{sec:amhjp_cr}

In this study, we investigate whether the cryptocurrency markets (Bitcoin and Ethereum) evolve over time, based on \citetapos{lo2004amh} AMH. Particularly, we estimate the degree of market efficiency based on a GLS-based time-varying model of \citet{ito2014ism,ito2016eme,ito2017aae}. The empirical results show that (1) cryptocurrency market efficiency varies with time, (2) the market efficiency of the BTC is higher than that of the Ethereum in most periods, and (3) the market has been evolving with high market liquidity. Therefore, we conclude that the empirical results support the AMH for the more established cryptocurrency market.

\clearpage

\section*{Acknowledgments}

The author would like to thank the Co-Editor, David Peel, an anonymous referee, Mikio Ito, Tatsuma Wada, and the conference participants at the 94th Annual Conference of the Western Economic Association International for their helpful comments and suggestions. The author is also grateful for the financial assistance provided by the Murata Science Foundation and the Japan Society for the Promotion of Science Grant in Aid for Scientific Research, under grant numbers 17K03809, 17K03863, 18K01734, and 19K13747. All data and programs used are available upon request.


\bigskip

\bigskip

\bigskip

\setcounter{table}{0}
\renewcommand{\thetable}{\arabic{table}}

\begin{table}[!htbp]
\caption{Descriptive Statistics and Unit Root Tests}
\label{crypto_table1}
\begin{center}\footnotesize
\begin{tabular}{c|cccc|cc|c}\hline\hline
 & Mean & SD & Min & Max &  ADF & Lag & $\mathcal{N}$ \\ \hline
 $R_{BTC}$ & 0.0018  & 0.0431  & -0.2662  & 0.3575  & -34.5442 & 1  & 2346  \\
 $R_{ETH}$ & 0.0028  & 0.0731  & -1.3021  & 0.4123  & -20.2283 & 2  & 1515  \\\hline\hline
\end{tabular}
\vspace*{5pt}
{
\begin{minipage}{320pt}
\scriptsize
{\underline{Notes:}}
\begin{itemize}
\item[(1)] ``ADF'' denotes the ADF test statistics and ``Lag'' denotes the lag order selected by the BIC.
\item[(2)] In computing the ADF test, a model with a time trend and constant is assumed. The critical value at the 1\% significance level for the ADF test is ``$-3.96$''.
\item[(3)] ``$\mathcal{N}$'' denotes the number of observations.
\item[(4)] R version 3.6.3 was used to compute the statistics.
\end{itemize}
\end{minipage}}%
\end{center}
\end{table}

\end{document}